\documentclass[a4paper]{spie}  
\usepackage[]{graphicx}
\usepackage{caption}
\usepackage{subfigure}
\usepackage{amsmath}
\usepackage{amssymb}
\usepackage{epsfig}
\usepackage{cite}
\usepackage{hyperref}
\usepackage{breakurl,color}
\title{WAHRSIS: A Low-cost, High-resolution Whole Sky Imager\\With Near-Infrared Capabilities}

\author{Soumyabrata Dev\supit{a}, Florian M. Savoy\supit{b}, Yee Hui Lee\supit{a}, Stefan Winkler\supit{b}
\skiplinehalf
\supit{a}School of Electrical and Electronic Engineering,\\Nanyang Technological University (NTU), Singapore 639798\\
\supit{b}Advanced Digital Sciences Center (ADSC),\\University of Illinois at Urbana-Champaign, Singapore 138632
}

\authorinfo{Send correspondence to YH Lee, E-mail: eyhlee@ntu.edu.sg}

\begin{document} 
  \maketitle 

\begin{abstract}
Cloud imaging using ground-based whole sky imagers is essential for a fine-grained understanding of the effects of cloud formations, which can be useful in many applications. Some such imagers are available commercially, but their cost is relatively high, and their flexibility is limited.  Therefore, we built a new daytime Whole Sky Imager (WSI) called Wide Angle High-Resolution Sky Imaging System. The strengths of our new design are its simplicity, low manufacturing cost and high resolution. Our imager captures the entire hemisphere in a single high-resolution picture via a digital camera using a fish-eye lens. The camera was modified to capture light across the visible as well as the near-infrared spectral ranges. This paper describes the design of the device as well as the geometric and radiometric calibration of the imaging system.
\end{abstract}



\section{INTRODUCTION}
\label{sec:intro}  

Whole sky imagers are becoming popular amongst the research community for a variety of applications and domains, such as aviation, weather prediction, and solar energy.  The resulting images are of higher resolution than what can be obtained from satellites, and the upwards pointing nature of the camera makes it easy to capture low-altitude clouds. They thus provide a useful complement to satellite images.

Our specific objective is to use such imagers in the analysis and prediction of signal attenuation due to clouds.  Ground-to-satellite or ground-to-air communication signals suffer from substantial attenuation in the atmosphere. Rain, clouds, atmospheric particles, and water vapor along the signal path affect the quality in various ways\cite{SiteDiversity,GammaDrop}. Analysis and prediction of those effects thus require accurate information about cloud formations along the communication path. An efficient data acquisition technique is required to detect and track the various types of clouds. Images captured with geo-stationary satellites are commonly used for this task. However those devices can only provide a limited spatial resolution. Such an approach generally fails in countries like Singapore, which has a small land mass, and where weather phenomenons are often very localized. Furthermore the downwards pointing nature of satellite images is a significant disadvantage for the capture of lower cloud layers\cite{Shields}.

Several models of automatic whole sky imagers have been developed at the Scripps Institute of Oceanography, University of California, San Diego, with cloud detection in mind~\cite{UCSD}. Yankee Environmental Systems commercialize the \emph{TSI-880}, which is used by many institutions \cite{Long,Souza}. Those imagers provide a good starting point for cloud analysis, but their design and fabrication is controlled by a company, making specialized adaptations hard or impossible. Besides, those devices are expensive (around 30,000-35,000~US\$ for the \emph{TSI-880}) and use very low resolution capturing devices.  Those limitations encouraged us to build our own.

In this paper, we propose a novel design of a whole sky imager called \emph{WAHRSIS} (Wide Angle High-Resolution Sky Imaging System). It uses mostly off-the-shelf components which are available on the market for an overall price of around 2,500~US\$. A simple overall design allows a home-made mounting of the various parts. The resulting images are of high resolution and can be used for various applications.

The paper is organized as follows: Section \ref{sec:design} presents the overall design of the device, including the various components and imaging system. Section \ref{sec:calibration} describes the geometric and radiometric calibration of the camera. Conclusions and future work are summarized in Section \ref{sec:conclusions}.

\section{WAHRSIS Design} \label{sec:design}


\subsection{Components} 
\label{sec:components}

\begin{figure}[htb]
\begin{center}
\includegraphics[width=0.35\textwidth]{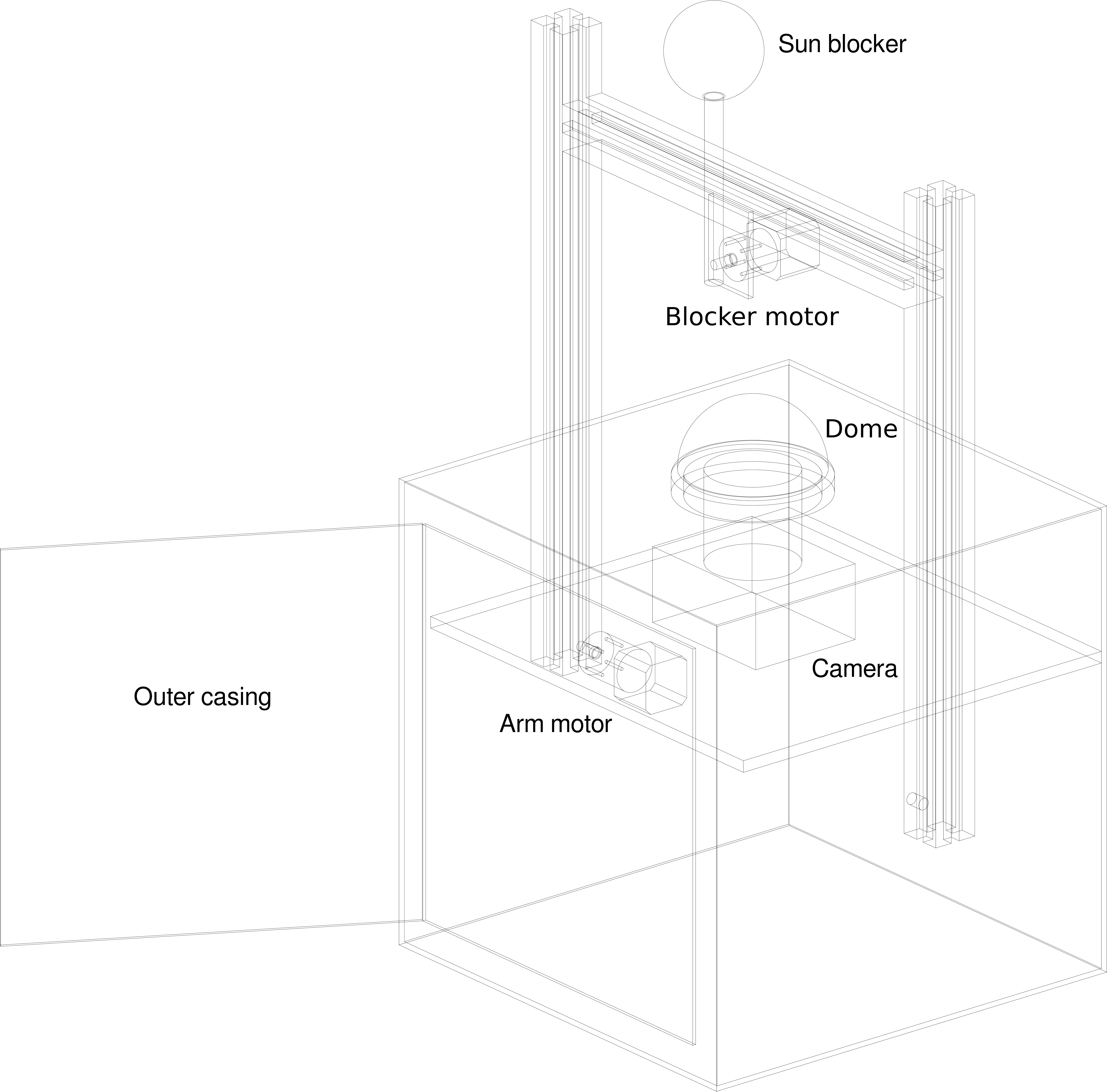}
 \caption{Drawing of \emph{WAHRSIS}.\label{fig:MechDesgn}}
\end{center}
\end{figure}

Figure \ref{fig:MechDesgn} shows a drawing of \emph{WAHRSIS}. Its main components are the following:
\begin{itemize}
\item The \emph{sun-blocker} occludes the light coming directly from the sun and thus reduces the glare in the circumsolar region of the captured image. It is composed of a main arm, which is rotated around the outside of the box by a motor. A polystyrene ball is fixed on top of a stem, which is attached to the main arm via a second motor rotating along the other axis.
\item An \emph{Arduino board} controls the two motors of the sun blocker. The algorithm is briefly explained in the Appendix \ref{append:motor}.
\item A \emph{built-in laptop} supervises the whole process. It controls the Arduino board through a serial communication port and the camera through a \emph{USB} connection and the Canon Digital Camera SDK. The captured images are stored on its hard drive and can be accessed via an Internet connection. As an alternative, a workstation situated outside the box can also be used.
\item \emph{Outer casing:} A hermetically-sealed box prevents moisture and dirt from affecting the internal components. It contains a transparent dome at the top in which the fish-eye lens of the camera is positioned.
\item The \emph{camera and lens} is discussed in more detail in Section \ref{sec:camera} below.
\end{itemize}

The device operates autonomously. However, some periodic manual maintenance interventions may be needed occasionally, for example to prevent the wear and tear of the motor gears. We have currently built one prototype of \emph{WAHRSIS}, which we placed on the roof-top of our university building. We plan to position several models across different parts of Singapore. Table \ref{cost} details all the components and their respective prices at the time of purchase \cite{FYP}.

\begin{table}[htb]
\centering
\begin{scriptsize}
    \begin{tabular}{ | l | r |}
    \hline
    \textbf{Items} & \textbf{Cost (in US\$)} \\ \hline
    Arduino Mega & 90 \\ \hline
    Arduino Mega 2560-control board	& 60 \\ \hline
    Bipolar 5.18:1 Planetary Gearbox Stepper & 25 \\ \hline
    Bipolar 99.55:1 Planetary Gearbox Stepper & 35 \\ \hline
    Stepper Motor Driver & 20 \\ \hline
    Real Time Clock & 15 \\ \hline
    Big Easy Stepper Motor Driver & 30 \\ \hline
    12 V DC Power Supply & 50 \\ \hline
    Base Hard plastic or Acrylic (Water-proof housing) & 120 \\ \hline
    Plastic Dome & 25 \\ \hline
    Metal Arm & 50 \\ \hline
    Sun Blocker & 5 \\ \hline
    Cables and accessories & 40 \\ \hline
    Sigma 4.5 mm F2.8 EX DC HSM Circular Fisheye Lens & 950\\ \hline
    Canon EOS Rebel T3i (600D) camera body & 360 \\ \hline
    Alteration for near-infrared sensitivity & 300 \\ \hline
    Built-in laptop & 350 \\ \hline
    \textbf{Total Cost} & \textbf{2525} \\ \hline
    \end{tabular}
    \caption{Cost Analysis for all the components of WAHRSIS}
    \label{cost}
    \end{scriptsize}
\end{table}

\subsection{Imaging System} \label{sec:camera}

The imaging system of \emph{WAHRSIS} consists of a \emph{Canon EOS Rebel T3i} (a.k.a.~\emph{EOS 600D}) camera body and a \emph{Sigma 4.5mm F2.8 EX DC HSM} Circular Fisheye Lens with a field of view of $180$ degrees. It captures images with a resolution of $5184 \times 3456$ pixels. Due to the lens design and sensor size, the entire scene is captured within a circle with a diameter of approximately $2950$ pixels (see Figure \ref{fig:wb}). 
A custom program running on the built-in laptop uses the \emph{Canon Digital Camera SDK} to control the camera. Images can be automatically captured at regular time intervals, and the camera settings can be adjusted as needed.

The sensor of a typical digital camera absorbs near-infrared light quite effectively, so much that there is an infra-red blocking filter behind the lens. We had this filter physically removed and replaced by a piece of glass, which passes all wavelengths.  The reasoning is that near-infrared light is less susceptible to attenuation due to haze\cite{sabine}, which is a common phenomenon in the atmosphere and especially around large cities. 

Haze consists of small particles in suspension in the air. Those have a scattering effect on the light whose incidence is modeled by Rayleigh's law, meaning that the intensity of the scattered light $E_s \sim E_0/\lambda^4$, where $E_0$ is the intensity of the emitted light, and $\lambda$ the wavelength of the light. We can thus reduce this effect by considering longer wavelengths, such as the ones in near-infrared. This model is however only valid for particles whose size is smaller than $\lambda/10$. Most cloud consist of larger particles; their resulting scattering effect obeys Mie's law, which is wavelength-independent.  For this reason, we expect our modified camera  to provide us sharper images of cloud formations.
    
\section{WAHRSIS Calibration} \label{sec:calibration}

The images captured by \emph{WAHRSIS} are of high resolution and capture the full sky hemisphere. However, the scene appears distorted by the fish-eye lens, and the colors are not rendered correctly due to the alteration for near-infrared sensitivity. Various calibration stages are thus required~\footnote{The source code of the various calibration processes for our sky camera WAHRSIS is available online at \url{https://github.com/Soumyabrata/WAHRSIS}.}. Section \ref{sec:white_balancing} discusses white balancing. Section \ref{sec:geometrical_calib} relates the geometric calibration due. Finally, Section \ref{sec:intensity} details the vignetting correction.

\subsection{White Balancing}\label{sec:white_balancing}
The camera of \emph{WAHRSIS} is also sensitive to near-infrared light and thus sees beyond the visible spectrum. Color calibration in the traditional sense is less meaningful for our camera, because it is sensitive to both visible and near-infrared light. However, white balancing is still necessary, as we will explain here.

\begin{figure}[htb]
\begin{center}

\subfigure[\emph{Auto}\label{fig:auto}]{\includegraphics[width=0.25\textwidth]{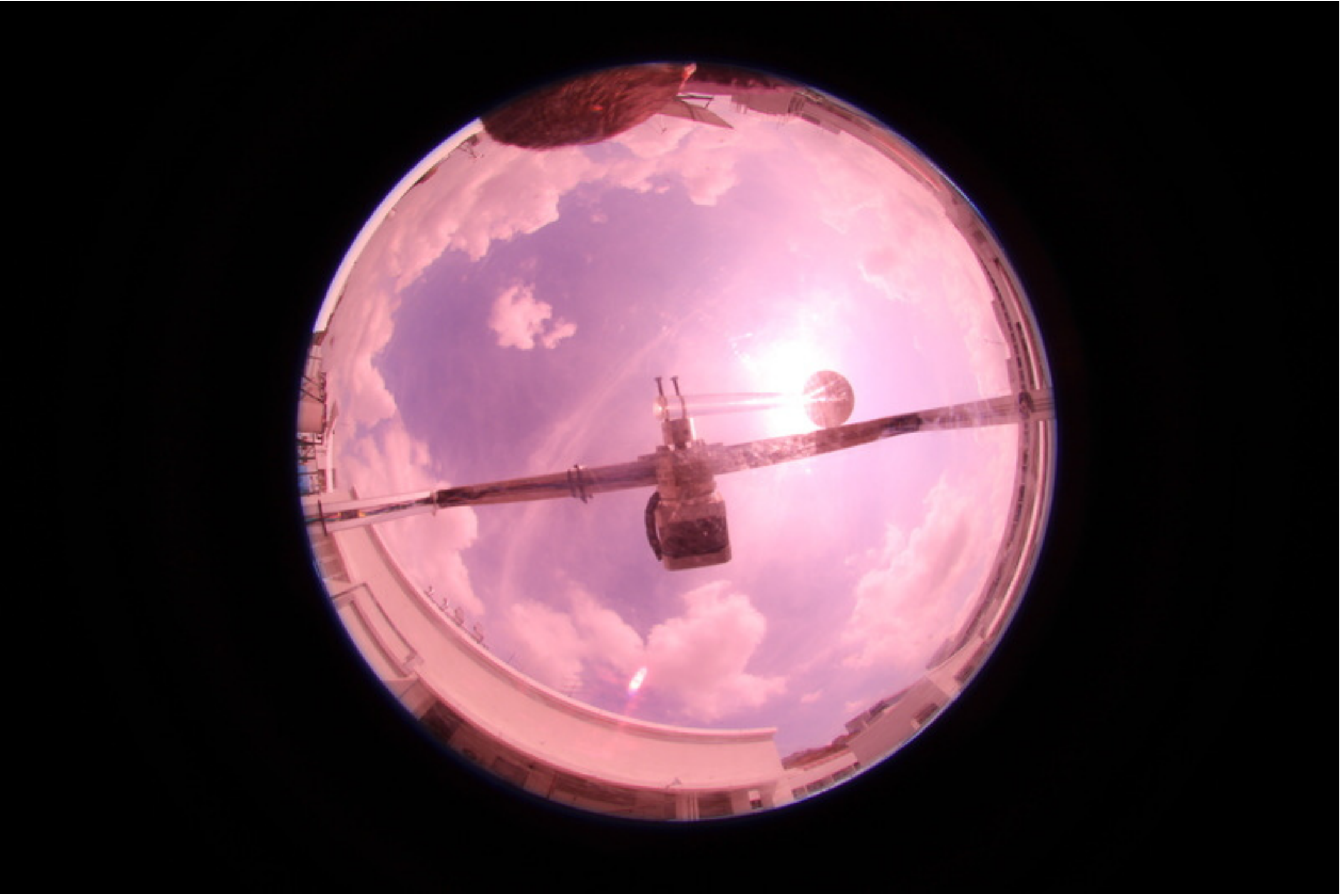}}
\subfigure[\emph{Custom}\label{fig:custom}]{\includegraphics[width=0.25\textwidth]{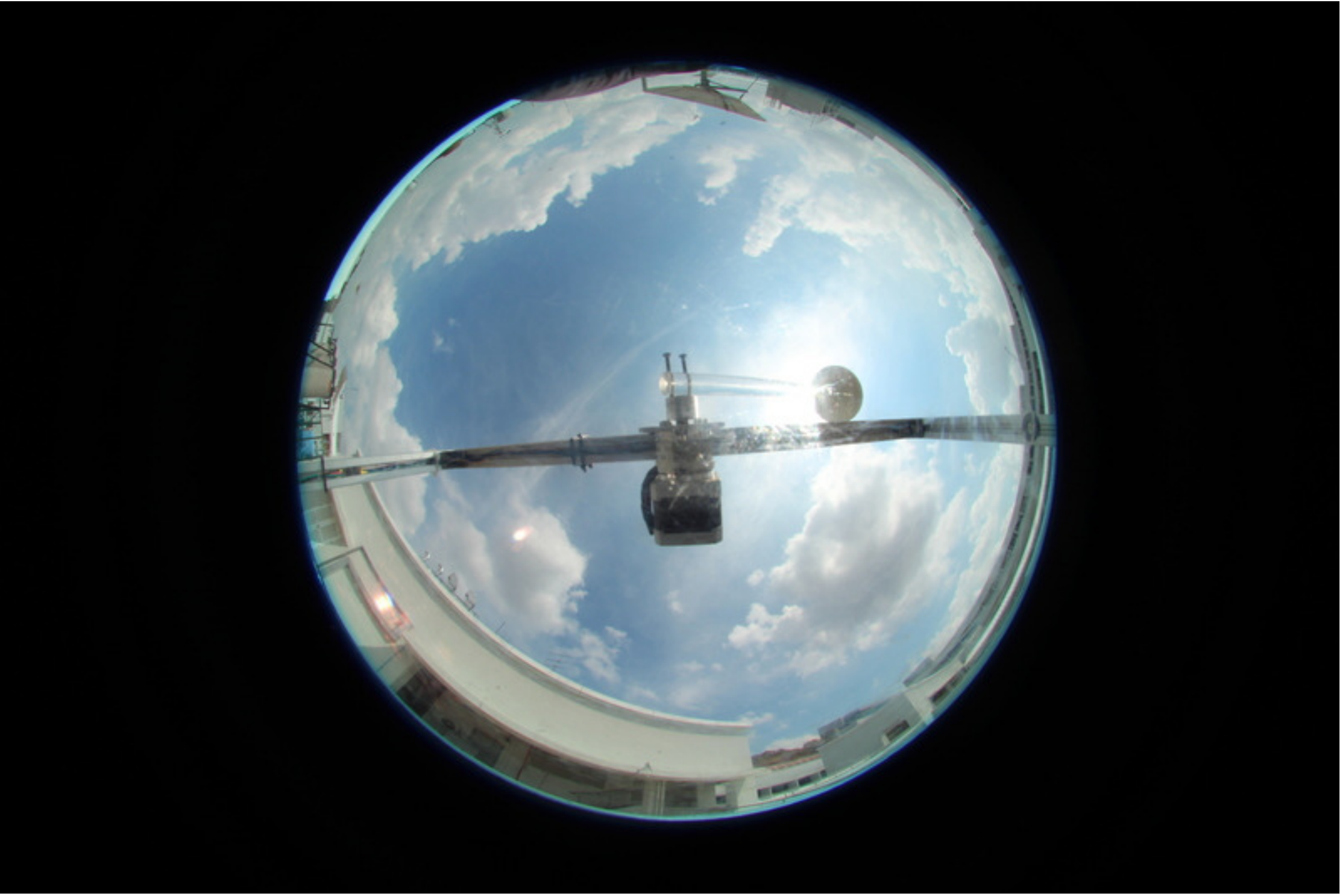}}
\caption{Captured images using automatic and custom white balance settings.}\label{fig:wb}
\end{center}
\end{figure}

The sensitivity of the various channels depends on the Bayer filter in front of the sensor. It is known that the red pixels show more near-infrared leakage than the blue or green ones. The automatic white balance settings were engineered for a non modified camera. As a result, images captured under an automatic white balance mode appear reddish, as shown in Figure \ref{fig:auto}. Fortunately, our camera also provides a custom mode. It relies on a known white patch in the scene, for which the camera computes the incident light intensity quantization parameters to render this patch white in the captured picture. The resulting image after custom white balancing looks visually plausible, as is shown in Figure \ref{fig:custom}.

Without white balancing, the red channel of our camera is prone to over-saturation and color clipping due to the additional near-infrared light reaching the sensor. The red channel of Figure \ref{fig:auto} contains 24.5\% of clipped values (i.e. where $R=255$), whereas the same channel in Figure \ref{fig:custom} contains only 0.7\% of them. This shows the importance of the custom white balancing, which manages to compensate quite well for the near-infrared alteration.

\subsection{Geometric Calibration} \label{sec:geometrical_calib}

A geometric calibration of the capturing device needs to be performed in order to obtain a precise model of the distortion introduced by the imaging system, which in our case includes the fish-eye lens and the dome\cite{HUOJuan}. This consists of determining the intrinsic parameters of the camera, which relates the pixel coordinates with their 3D correspondences in a camera coordinate system. On the other hand, extrinsic parameters express the translations and rotations required to use other 3D world coordinate systems, such as the ones span by the checkerboard borders. Intrinsic parameters give all the information needed to relate each pixel of a resulting image with the azimuth and elevation angles of the incident light ray and vice-versa, which is essential to be able to measure the physical extent of clouds.

\subsubsection{Fisheye Calibration Background} \label{sec:rel_studies}

\begin{figure}[htb]
\begin{center}
\includegraphics[width=0.4\textwidth]{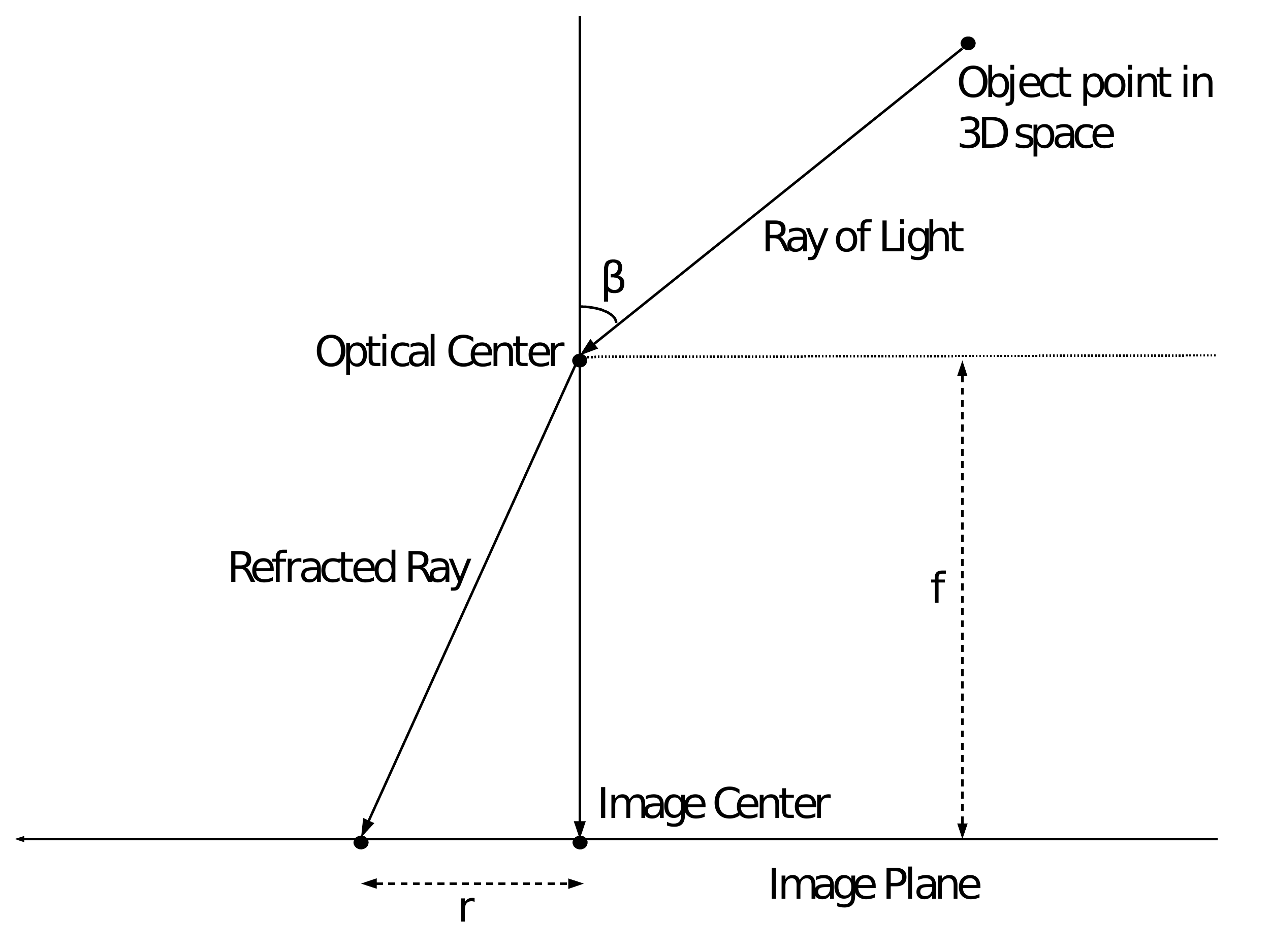}
\caption{Schematic representation of object space and image plane\label{fig:Fig6}}
\end{center}
\end{figure}

The modeling of a typical (non-fisheye) camera uses the pinhole model. However, this is not applicable to our case because of the very wide capturing angle. A schematic representation of the refraction of an incident ray in a fisheye lens is shown in Figure \ref{fig:Fig6}. Most common approaches relate the incident angle ($\beta$) with the radius on the image plane ($r$), assuming that the process is independent of the azimuth angle.

Several theoretical models relating those two values exist:
\begin{itemize}
\item Stereographic projection: $r=2f\tan(\beta/2)$
\item Equidistance projection: $r=f\beta$
\item Equisolid angle projection: $r=2f\sin(\beta/2)$
\item Orthogonal projection: $r=f\sin(\beta)$
\item Perspective projection: $r=f\tan(\beta)$
\end{itemize}
Fisheye lens manufacturers usually claim to obey one of those, but those equations cannot be exactly matched in practice. However, they can be approximated by a polynomial due to the Taylor series theory. Some of the calibration techniques make use of this, as it easily allows a slight deviation from the projection equation. Two polynomial functions can also be used to cope with possible radial and tangential distortions caused by the lens. 

Most calibration techniques use a checker-board with a known pattern. Shah et al.\cite{shah1996intrinsic} introduce a calibration method where such a polynomial is used. Furthermore, they model radial and tangential distortions. The optical center of the camera has to be found using a low power laser beam, which is cumbersome in practice. Bakstein et al.\cite{bakstein2002panoramic} use a spherical retina model, but do not provide any extrinsic parameter estimation technique and use cylindrical patterns. Sturm et al.\cite{sturm2004generic} introduce a very generic model, where a camera is described by the coordinates of a set of rays and a mapping between those and the image pixels. However the authors experienced difficulties with a fish-eye lens, especially on the side view.  Kannala et al.\cite{Kannala_calib} use polynomials to describe the calibration and the radial and tangential distortions, but they also assume that the positions of the pattern on the checker-board are known.  Finally, Scaramuzza et al.\cite{scaramuzza2006toolbox} provide an algorithm estimating both intrinsic camera parameters and the extrinsic parameters describing the positions of the checker-board. They provide a full \emph{MATLAB} implementation of their method, making it the easiest to use.  We have chosen to use this method and enhanced it for our needs.

\subsubsection{Calibration Method} \label{sec:calibration_method}

The calibration toolbox proposed by Scaramuzza and al.\cite{scaramuzza2006toolbox} takes as input a number of images containing a known checker-board pattern. The first step is corner detection, which is performed by the toolbox. User interaction is required when the automatic detection algorithm fails. Those points are then used to find the parameters by a least-squares linear minimization, followed by a non-linear refinement with a maximum likelihood criterion. The toolbox uses a polynomial to model the ray refraction as well as a $2\times 2$ matrix transformation and a translation to cope with small axis misalignment and the digitizing process. The intrinsic parameters thus consist of both the coefficients of the polynomial as well as the values of the matrix and the translation parameters, which describe the position of the center of the image.

\begin{figure}[htb]
\begin{center}
\includegraphics[width=0.2\textwidth]{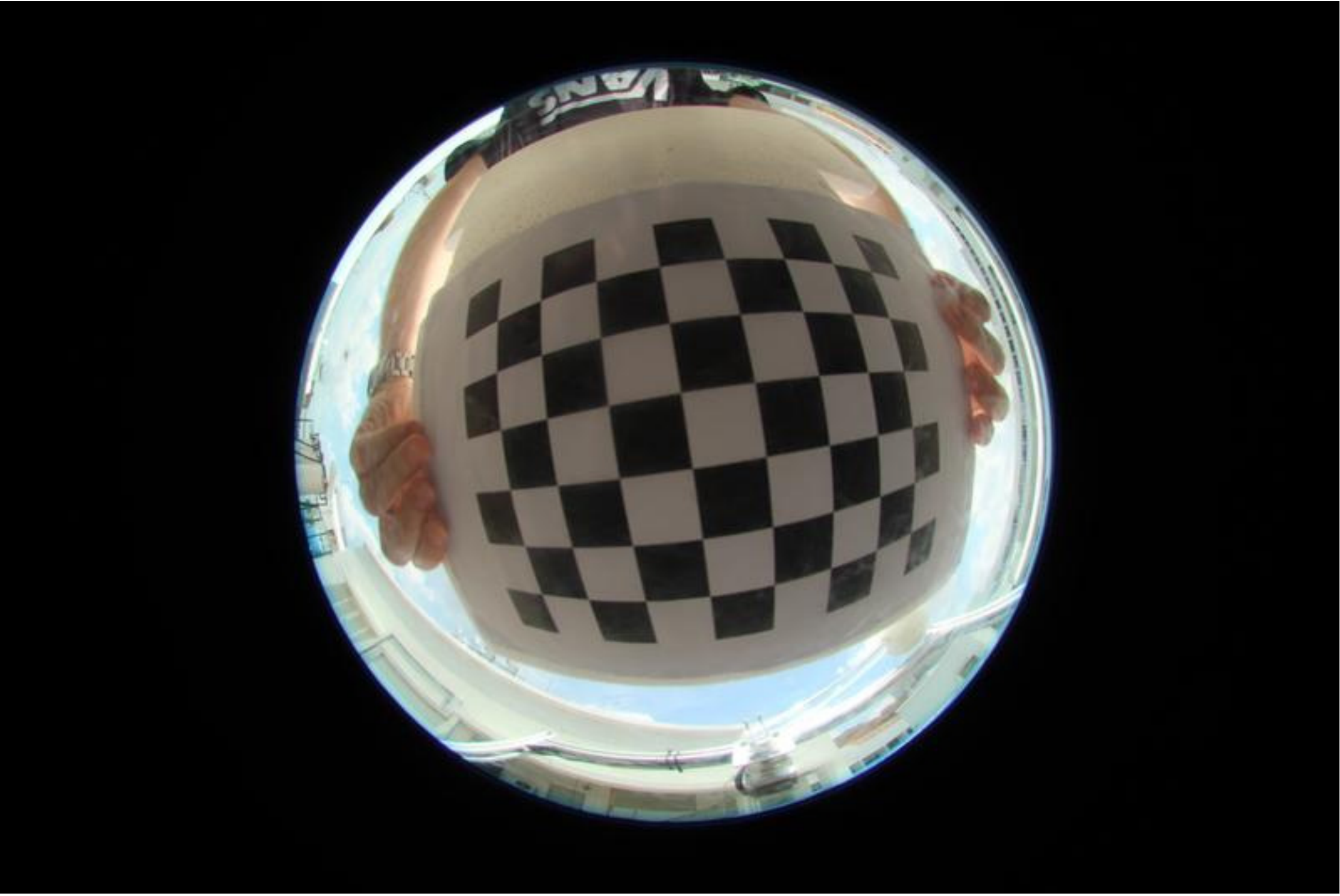}
\includegraphics[width=0.2\textwidth]{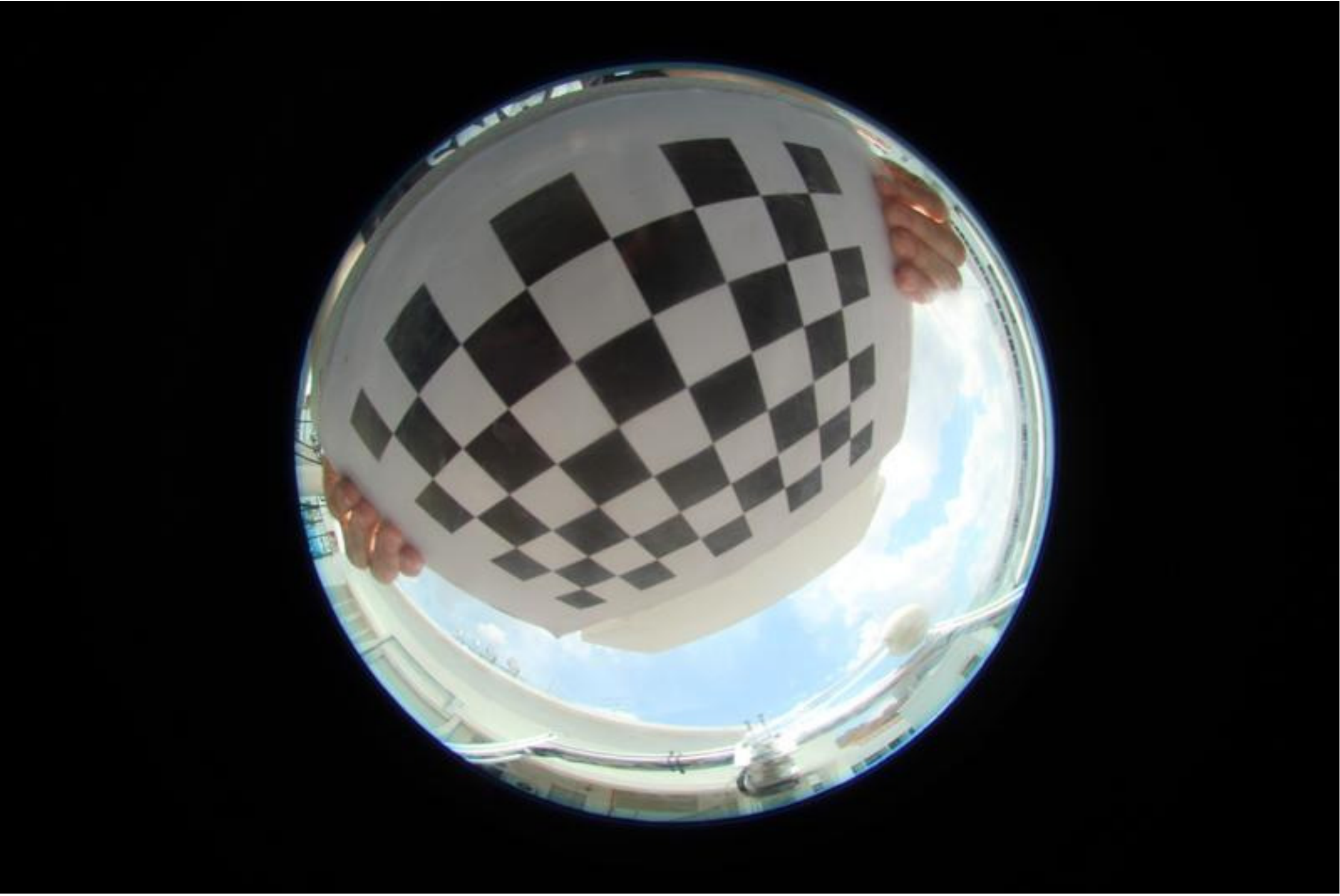}
\includegraphics[width=0.2\textwidth]{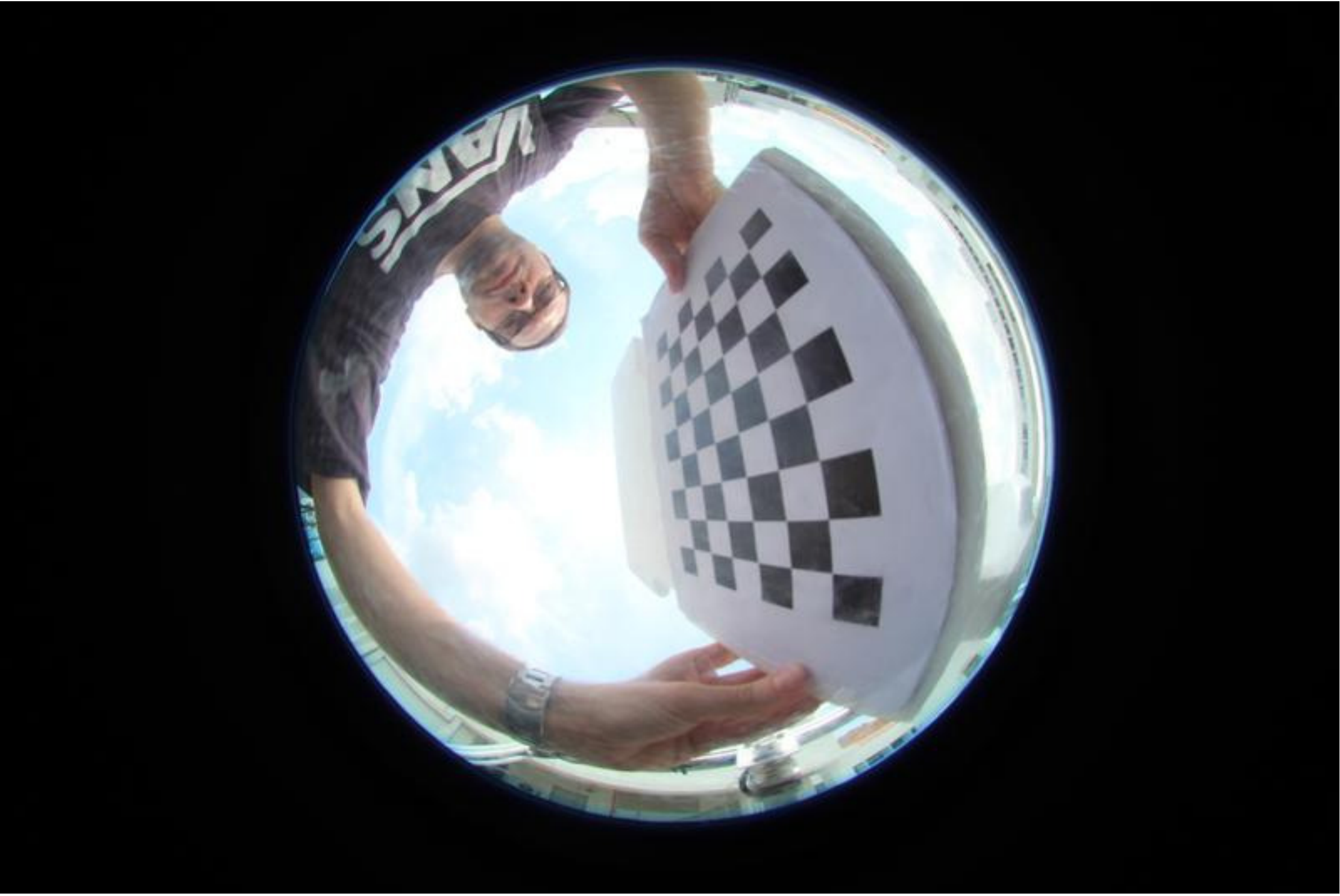}
\includegraphics[width=0.2\textwidth]{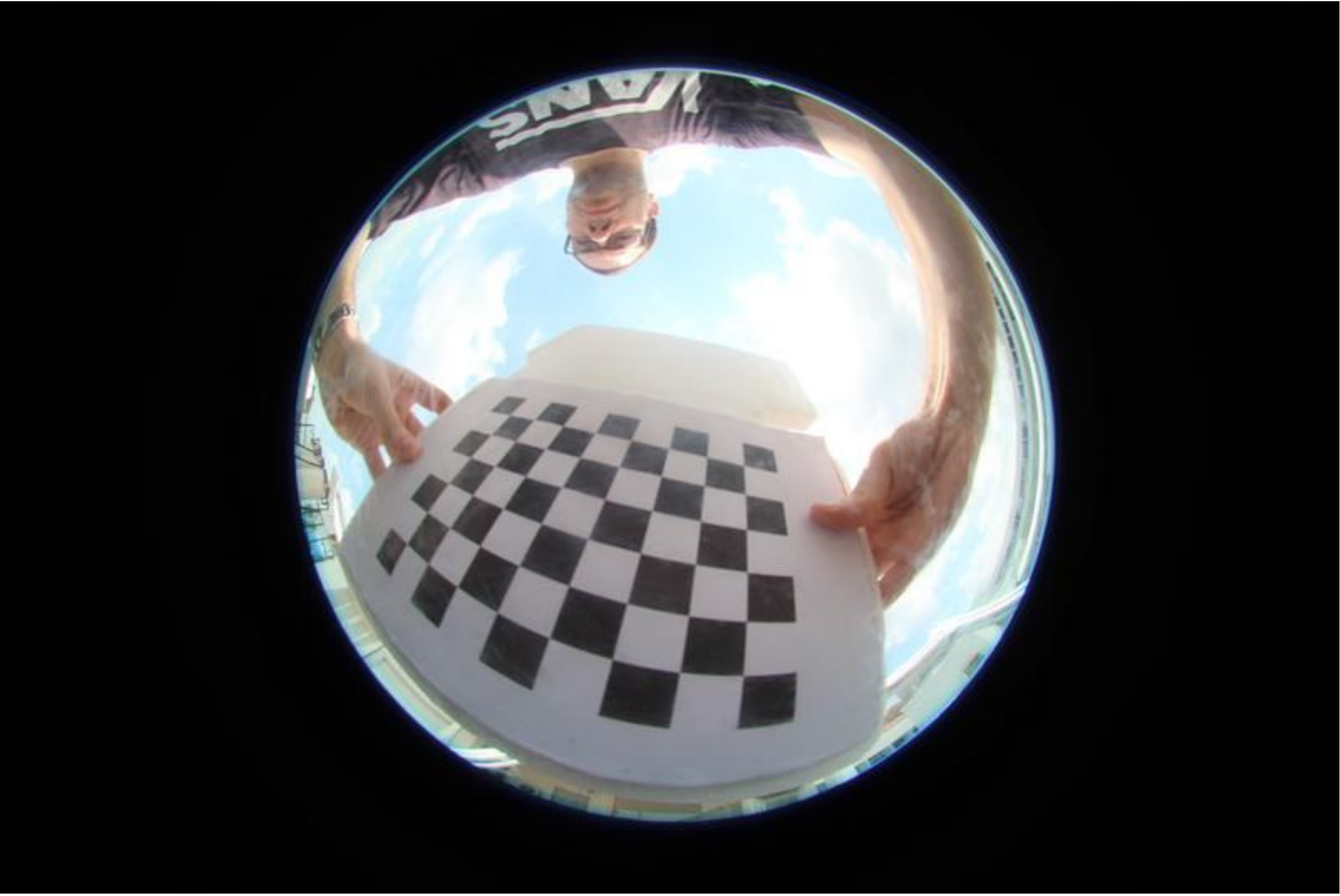}
\caption{Sample of the checker-board images used for the calibration\label{fig:calibration_images}}
\end{center}
\end{figure}

Figure \ref{fig:calibration_images} shows a sample of the images we used for the calibration.

The toolbox also outputs the re-projection errors, i.e. the difference between the original point location computed during the corner detection and the one resulting from the estimated parameters. This is done on the same images as the ones used for the parameter estimation. However we noticed that the returned errors do not necessarily relate the quality of the calibration, as the re-projection sometimes looks more accurate than the corner detection or the user input in estimating the coordinates of a grid corner. Furthermore we would like to estimate a general error which does not only rely on the location of the checker-boards in the images used for the fitting. We also observed that the choice of the input images has a significant influence on the output of the algorithm. We thus modified the calibration process in the following way: we randomly split our captured checker-board images into a training set of 10 images and a validation set of 6 images. We compute the parameters using the training set and the re-projection error using the validation set. We repeat this process 50 times with different randomized splits and finally consider the parameters leading to the smallest error on the validation set as the most accurate one. In this way we can reduce the uncertainty of the whole process. 

Our fish-eye lens (\emph{Sigma 4.5mm F2.8 EX DC HSM}) was designed by the manufacturer to follow the equisolid projection for visible light, but may be less accurate for the near-infrared spectrum. There is also a transparent dome on top of the camera lens, which may have an additional refracting effect on the incident rays. The toolbox models the calibration function as a polynomial and can thus incorporate these effects when fitting the coefficients.  We have set the maximum degree of the polynomial to 4, as advised by the authors of the toolbox. We indeed noticed a bigger re-projection error with smaller degrees. However when the maximum degree is set to a higher number, the  values of the coefficients of the degrees higher than 4 are almost zero.

\begin{figure}[htb]
\begin{center}
\subfigure[Relationship between the incident angle of a light ray and the corresponding distance from the image center.\label{fig:calibration_results}]{\includegraphics[width=0.48\textwidth]{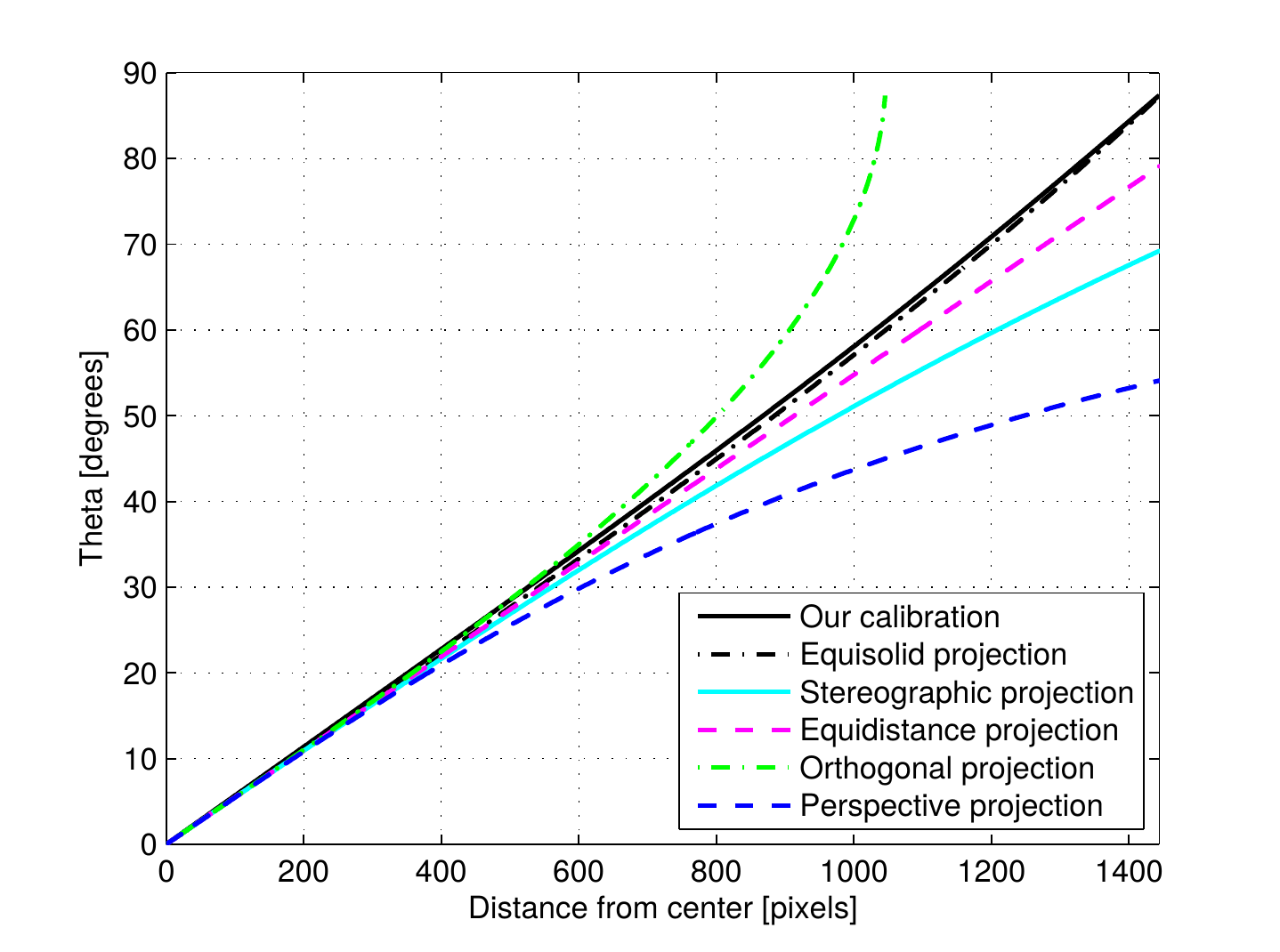}}\quad
\subfigure[Difference between equisolid model and our imaging system.\label{fig:diff_model_calib}]{\includegraphics[width=0.48\textwidth]{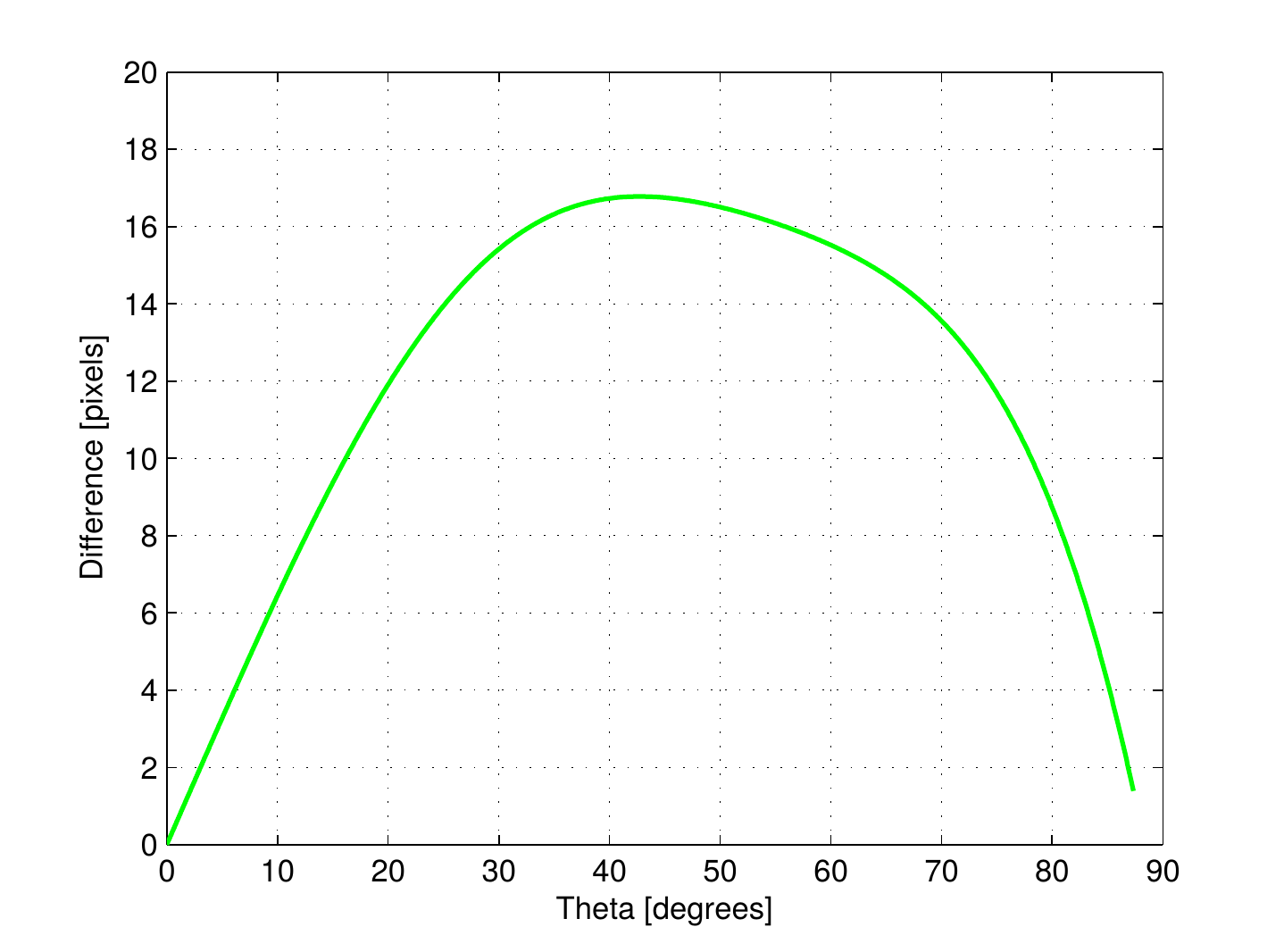}}
\caption{Result of the calibration process compared to the various theoretical projection models.}
\end{center}
\end{figure}

Figure \ref{fig:calibration_results} shows the result of the resulting calibration compared to the various theoretical projection models. Figure \ref{fig:diff_model_calib} shows the distance in pixels between our estimated calibration parameters and the various projection models. We see that we reach a distance of more than 16 pixels for angles around 40 degrees. We also find that the maximum viewing angle of the whole device is 178.7 degrees, slightly less than hemispheric.

\begin{figure}[htb]
\begin{center}
\includegraphics[width=0.48\textwidth]{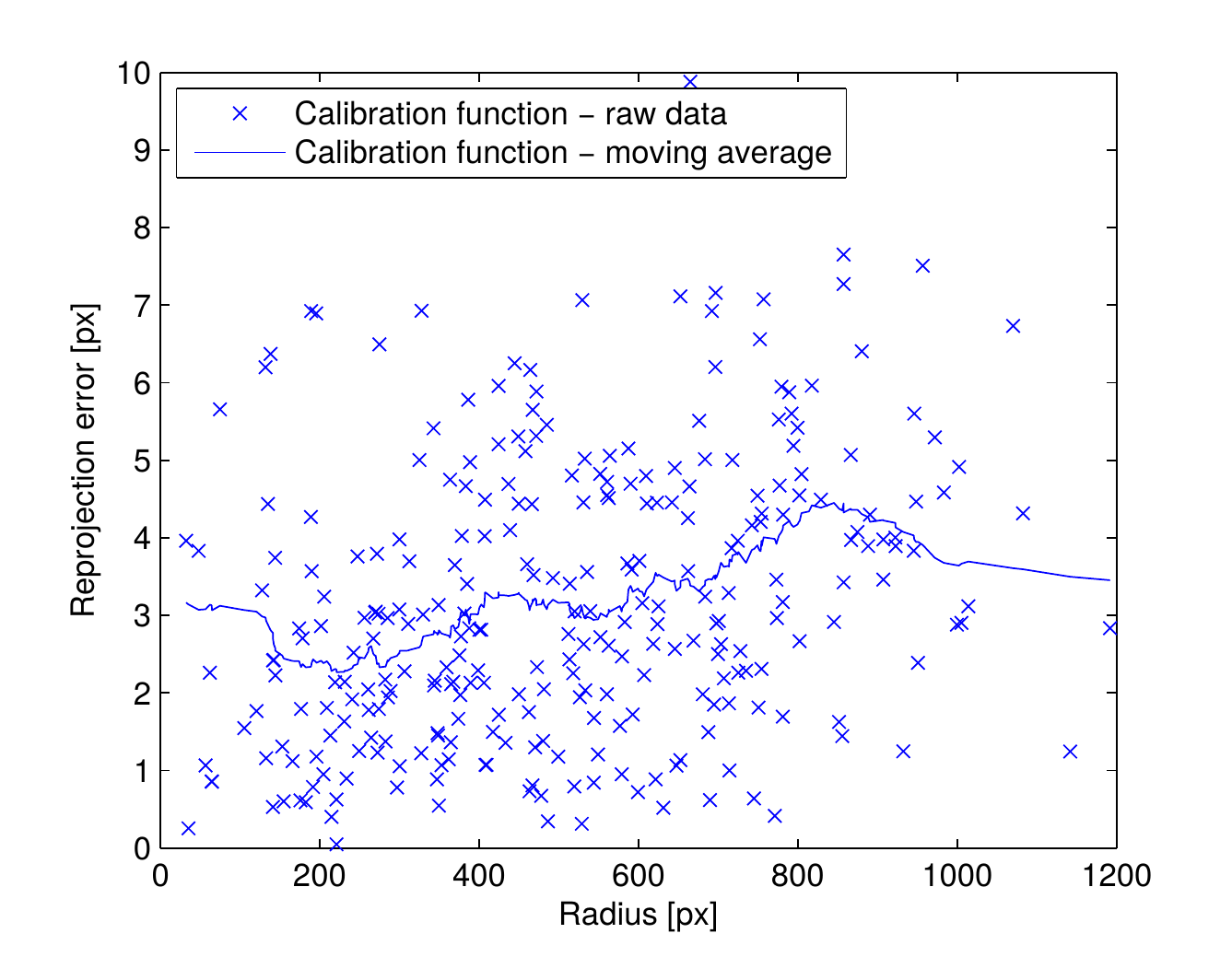}
\caption{Re-projection errors as a function of the distance to the image center (crosses: raw data; line: moving average).\label{fig:reprojection_error_bw}}
\end{center}
\end{figure}

Figure \ref{fig:reprojection_error_bw} shows the re-projection error for all the points of the grids in the validation set. A moving average with a window of 50 pixels is also shown. The average re-projection error for our model is 3.87 pixels. We can see a minor increase of the error with distance from the image center (and thus the elevation angle of the incident ray).

\newpage
\subsubsection{Chromatic Aberration} \label{sec:chromatic_aberration}

\begin{figure}[htb]
\begin{center}
\includegraphics[width=0.21\textwidth]{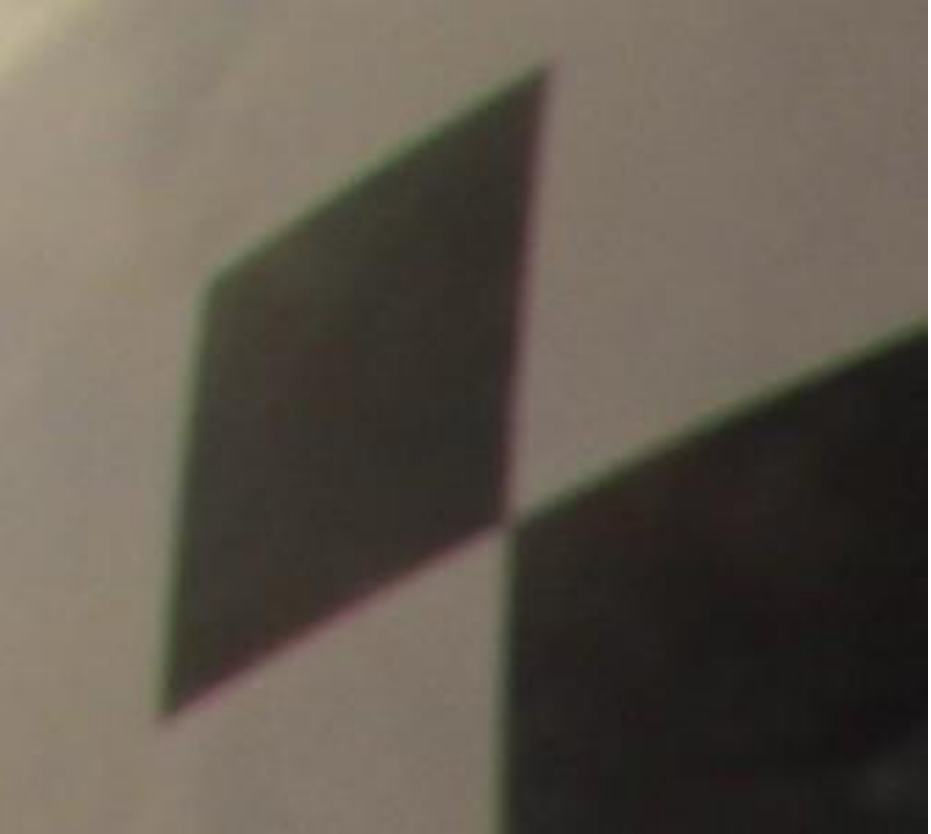}
\caption{Example of chromatic aberration. Notice the green and magenta artifacts at the edges of the squares.\label{fig:example_chrom_aberr}}
\end{center}
\end{figure}

The refractive index of lenses is wavelength-dependent. This means that light rays from different parts of the spectrum converge at different points on the sensor. Chromatic aberration refers to the color distortions introduced to the image as a result this phenomenon. Since the camera of \emph{WAHRSIS} is sensitive to near-infrared, a larger part of the light spectrum is captured by the sensor, making the device more prone to this issue. Figure \ref{fig:example_chrom_aberr} shows chromatic aberration on an image captured by the camera.

\begin{figure}[h!]
\begin{center}
\includegraphics[width=0.48\textwidth]{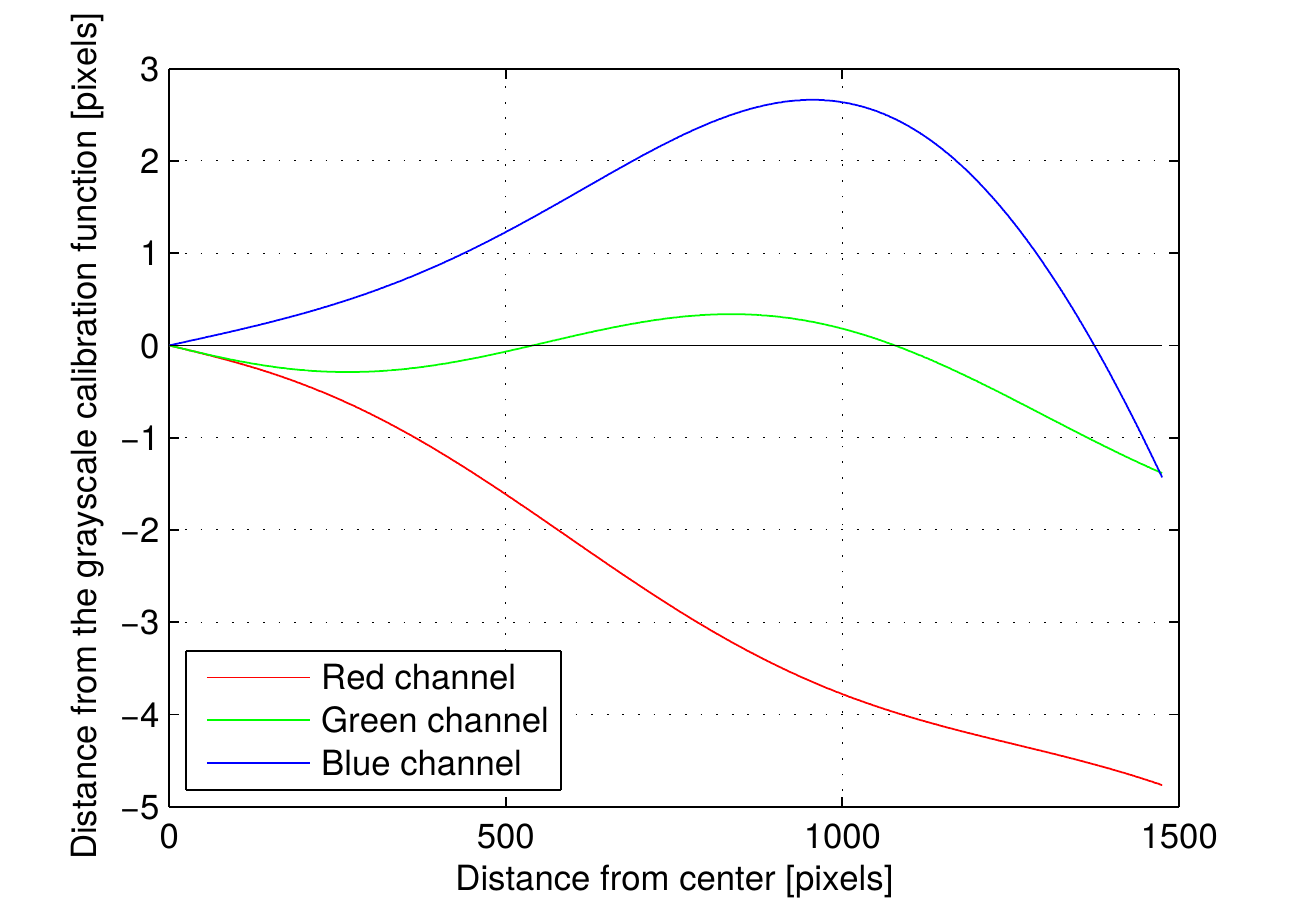}
\caption{Difference between the calibration function of each color channel and the grayscale calibration function.\label{fig:difference_RGB}}
\end{center}
\end{figure}

\begin{figure}[htb]
\begin{center}
\subfigure[Grayscale calibration.\label{fig:projBw}]{\includegraphics[width=0.21\textwidth]{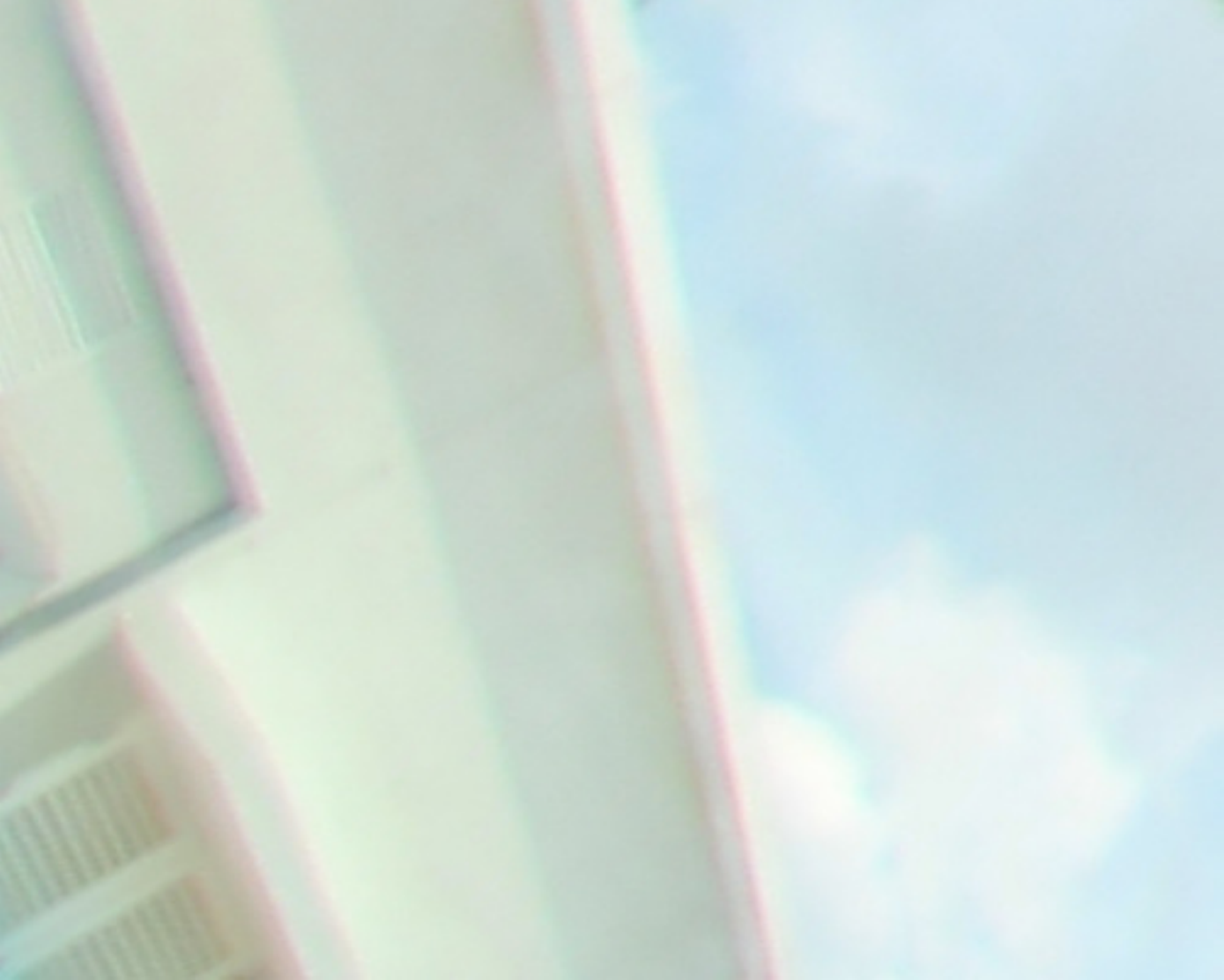}} \quad
\subfigure[Color-channel-specific calibration.\label{fig:projDiffChannels}]{\includegraphics[width=0.21\textwidth]{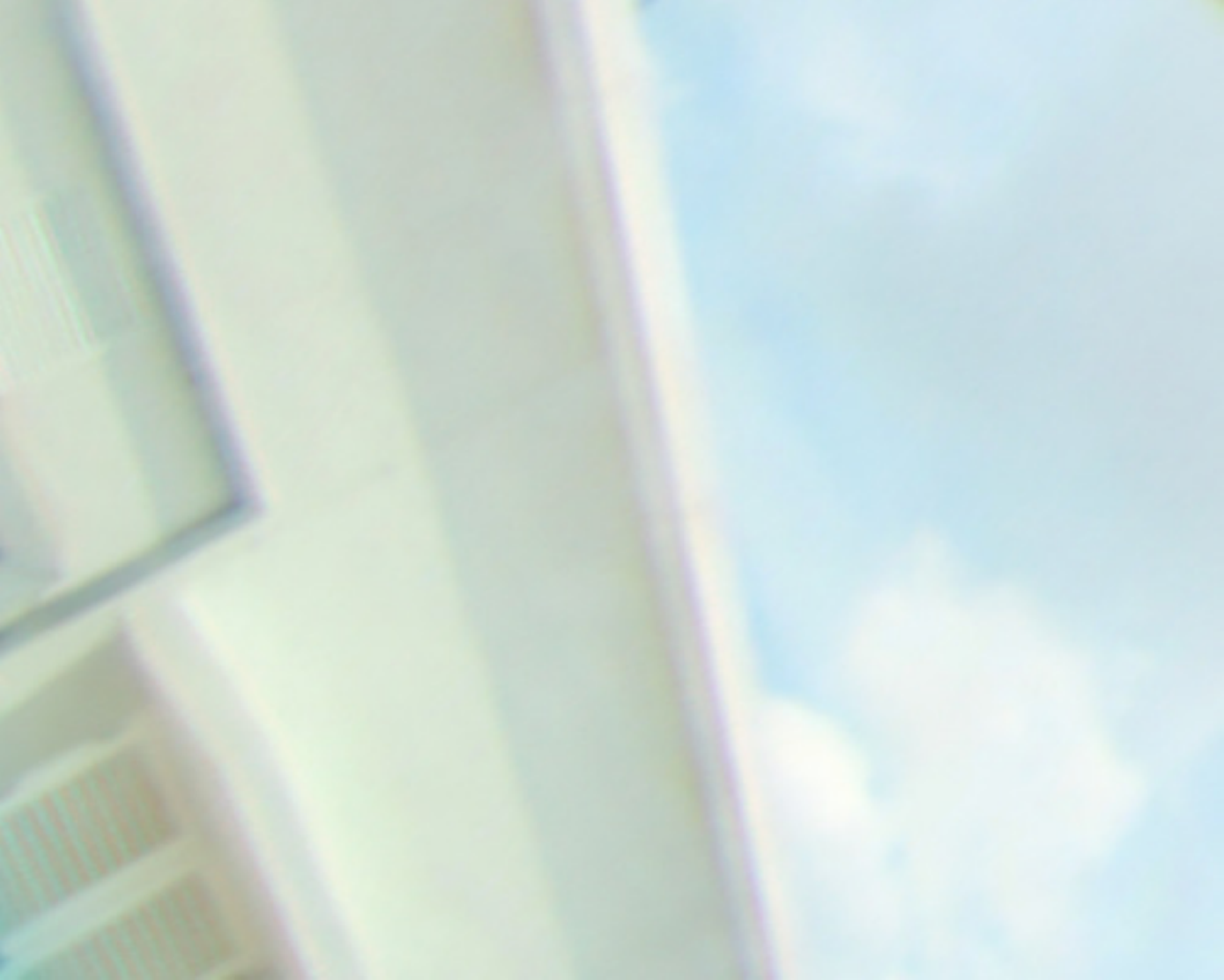}}
\caption{Section of an image taken with \emph{WAHRSIS} that was rectified using grayscale (left) and color-channel-specific (right) calibration. The chromatic aberration in the right image is much less noticeable.}
\label{fig:proj}
\end{center}
\end{figure}

The calibration process described in the previous section uses grayscale images. In order to obtain more accurate calibration functions, we apply the same process on each of the red, green, and blue channels individually. As shown in Figure \ref{fig:difference_RGB}, we observe small differences in the resulting calibration functions, especially for the red channel.  Visually, the difference is quite striking, as can be seen in Figure \ref{fig:proj}.

\subsection{Vignetting Correction}\label{sec:intensity}

Vignetting refers to a darkening toward the corners of the image which is introduced during the capturing process. It has several origins. Natural vignetting is due to the incident light rays reaching the camera sensors with varying angles. It is commonly modelled by the cosine fourth law. The camera of \emph{WAHRSIS} is particularly prone to this effect due to its wide angle. Optical vignetting describes the shading caused by the lens cylinder itself, blocking part of the off-axis incident light. This is aperture-dependant.\cite{toothwalker}

\begin{figure}[htb]
\begin{center}
\includegraphics[width=0.45\textwidth]{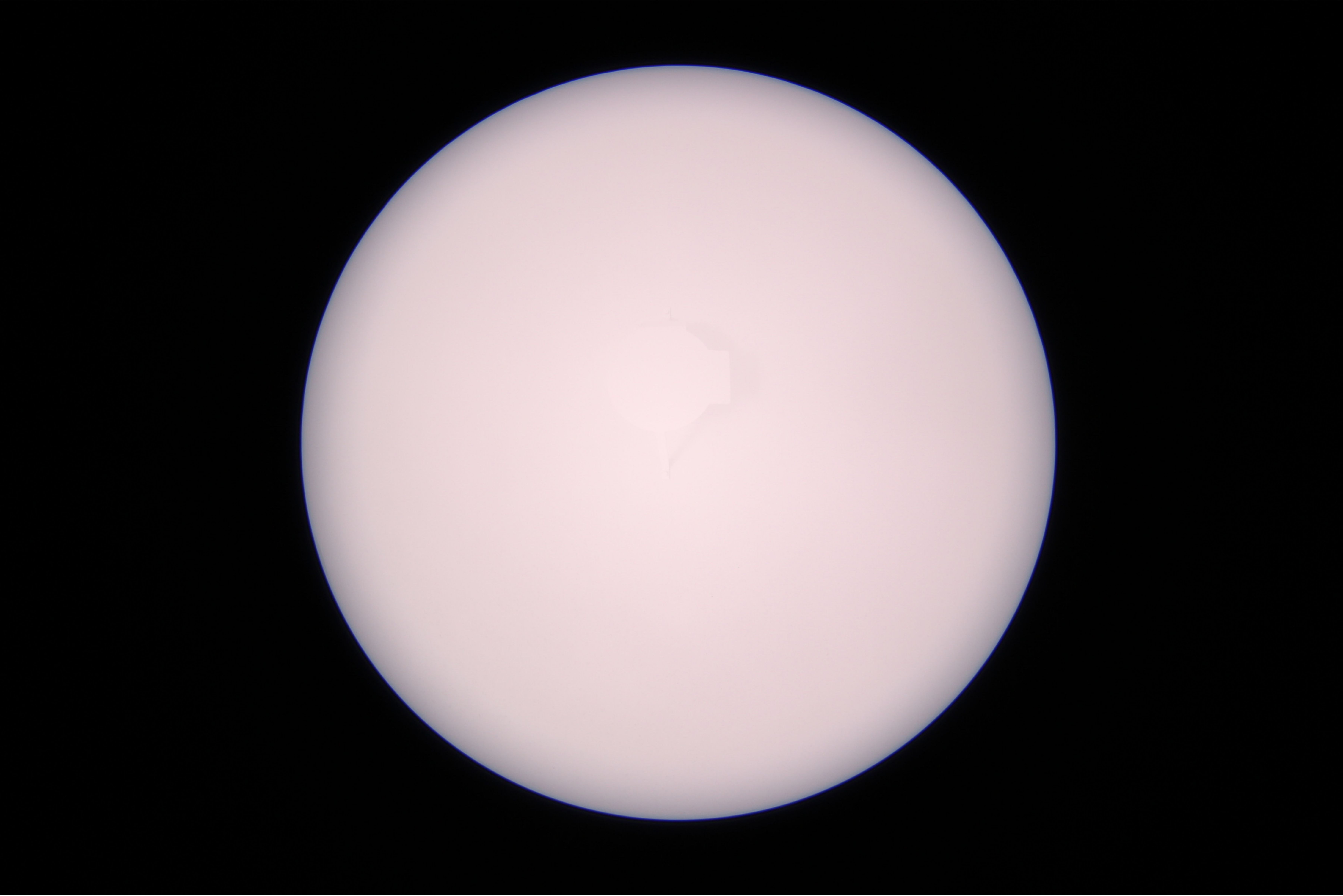}
\caption{Image captured inside the integrating sphere\label{fig:sphere}}
\end{center}
\end{figure}

We use an integrating sphere to analyze this phenomenon. It consists of a large sphere, whose interior surface is made of a diffuse white reflective coating, resulting in a uniform light source over the whole surface. A LED light source as well as the camera are placed inside. Figure \ref{fig:sphere} shows an image captured inside this sphere. Notice the higher brightness at the center. This image has been taken with the biggest available aperture ($f/2.8$) and constitutes the worst case with regards to vignetting. Smaller aperture values will result in less significant distortions.

\begin{figure}[htb]
\begin{center}
\subfigure[Luminance as a function of the distance from the image center\label{fig:pro_distance}]{\includegraphics[width=0.48\textwidth]{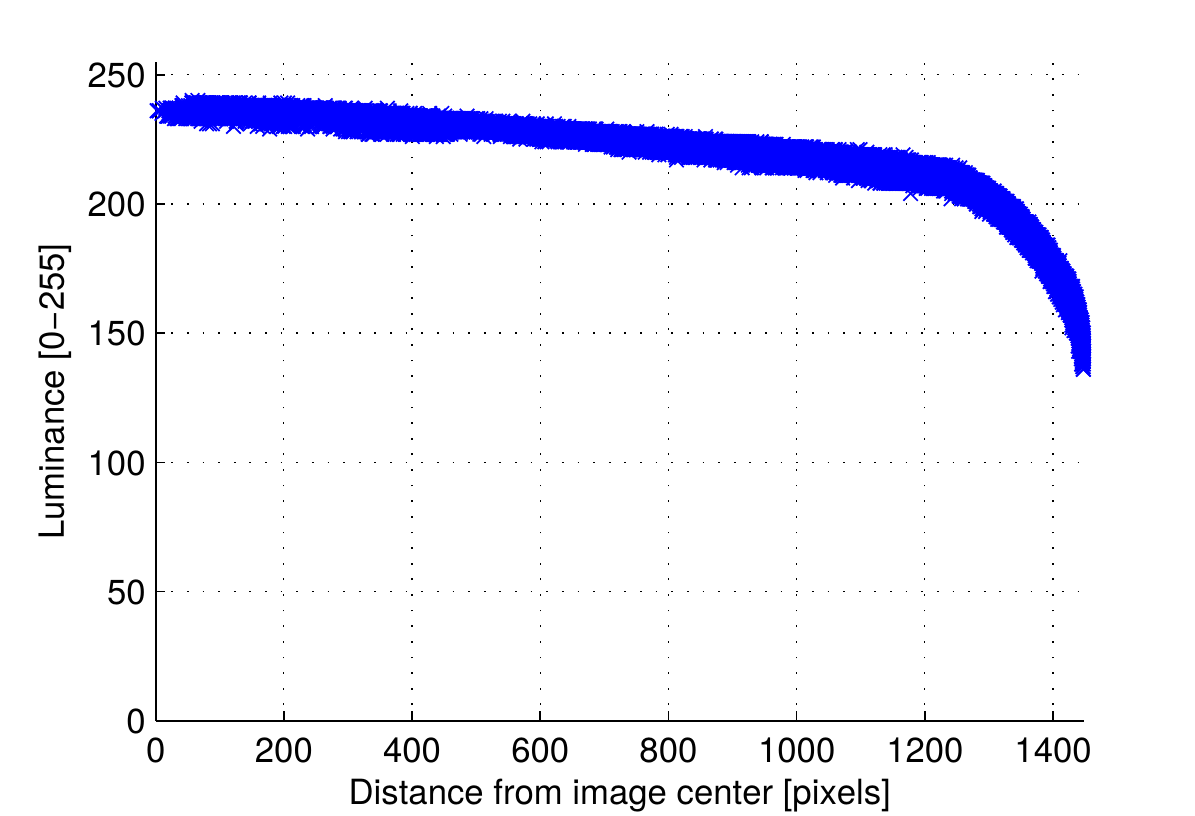}}\quad
\subfigure[Correction coefficients in red and scatter plot of all the values in blue. \label{fig:coefficients}]{\includegraphics[width=0.48\textwidth]{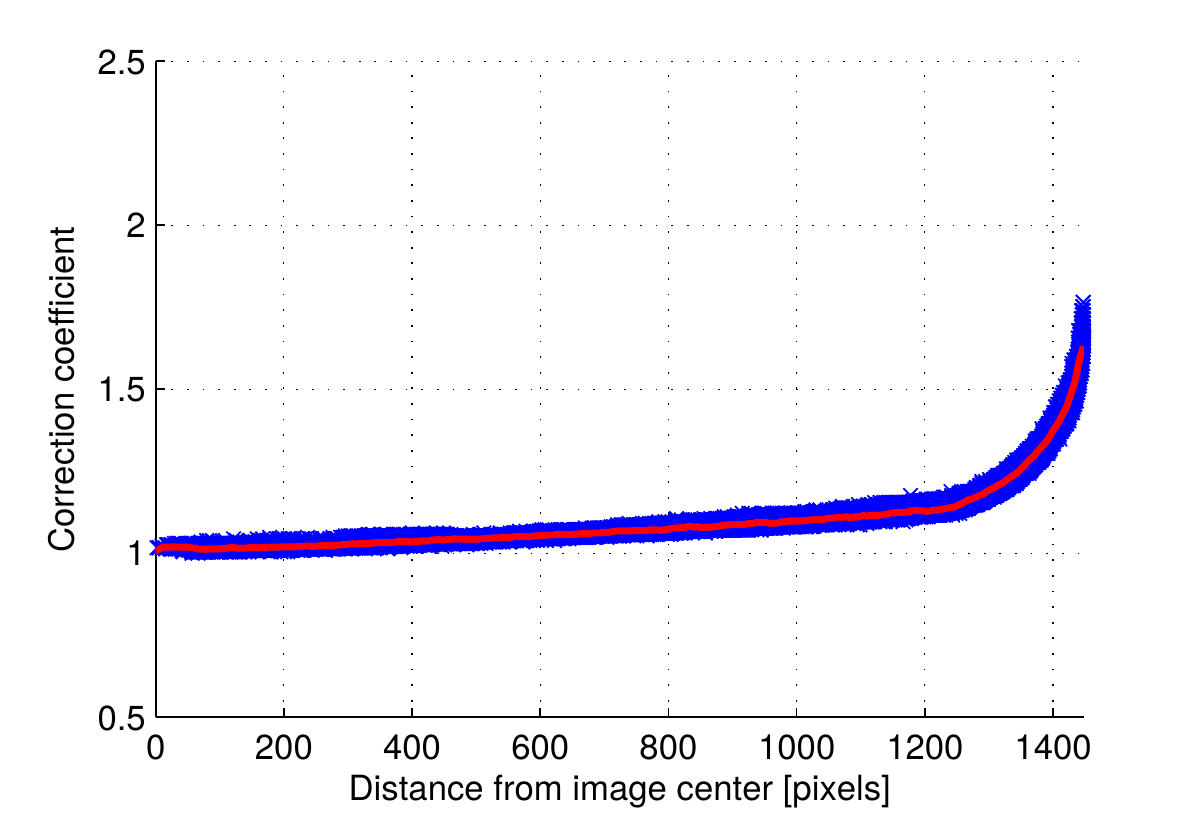}}
\caption{Computation of the vignetting correction coefficients}
\end{center}
\end{figure}

Figure \ref{fig:pro_distance} shows the luminance of the pixels inside the lens circle as a function of the distance from the image center. We normalize those values and then take their inverse. We fit a moving average and use this result as a radius dependent correction coefficient (Figure \ref{fig:coefficients}). We repeat this process for each aperture setting.

\section{Conclusion} \label{sec:conclusions}

We have presented WAHRSIS, a new whole sky imager. Its advantages are low cost and simple design. We described its components and the calibration of its imaging system. It provides high resolution images of the whole sky hemisphere, which can be useful for many various applications.

In our future work we will use the resulting images for detailed cloud analysis. We plan to deploy several devices across the country to estimate cloud cover, cloud bottom altitude, cloud movement, etc. We will also investigate the effects of the near-infrared capabilities for these analyses in more detail. Finally, we are working on an improved version of \emph{WAHRSIS} with a better sunblocker design.

\appendix
\section{Sun Blocker Positioning} \label{append:motor}
\begin{figure}[htb]
\begin{center}

\subfigure[Side and top views]{\includegraphics[clip=true, trim={0cm 0cm 0cm 0cm},height=5.5cm]{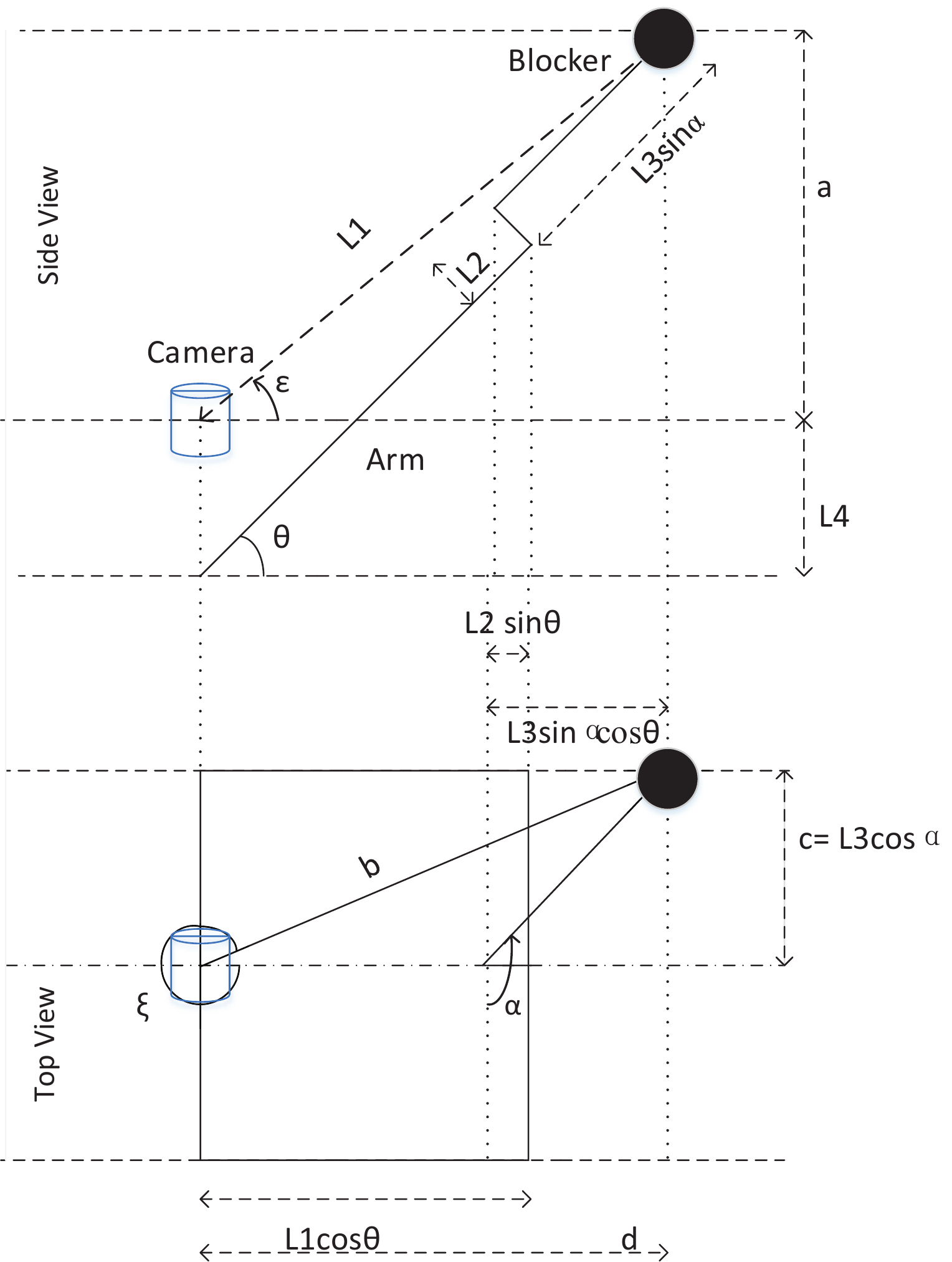}}\quad
\subfigure[3D view]{\includegraphics[clip=true,  width=0.35\textwidth]{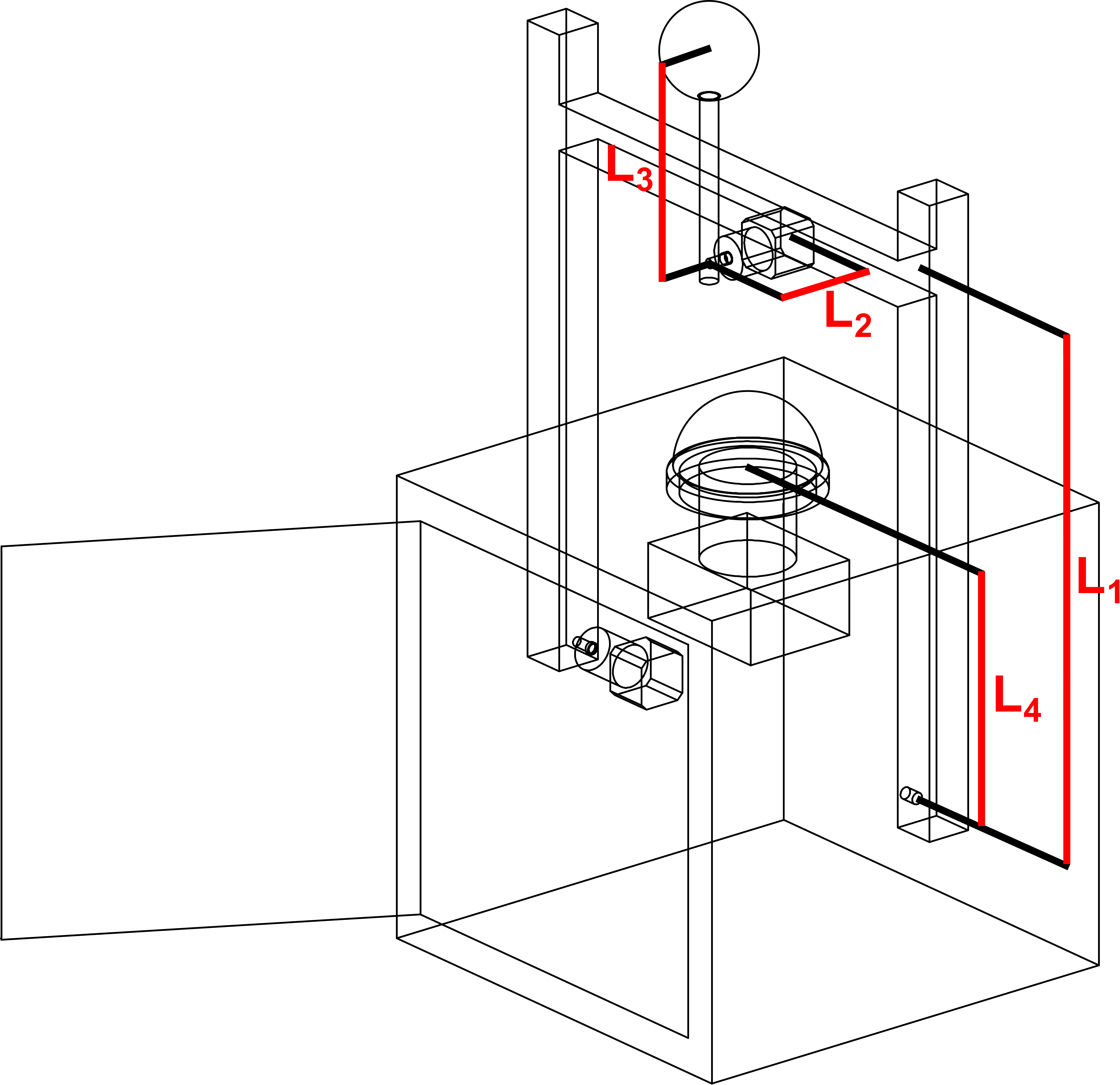}}

\caption{Representation of WAHRSIS with the lengths and angles required for the computation of the motor angles.\label{fig:SideTopWSI_sub}}
\end{center}
\end{figure}

In order to block the light coming directly from the sun, the sun blocker head should intersect the ray from the sun to the camera. We use the Jean Meeus algorithm\cite{SPA} to compute the azimuth and elevation angles of this ray, which we call $\xi$ and $\varepsilon$ respectively. The algorithm has an error of $\pm 3\cdot 10^{-4}$ degrees. The various lengths and angles required for the computations are shown in figure \ref{fig:SideTopWSI_sub}. $L_{1}$ represents the length of the main arm. $L_{2}$ is the distance of the motor shift between the blocker and the arm.  $L_{3}$ denotes the length of the stem under the blocker head. $L_{4}$ denotes the length between the camera center and the arm motor axis. The two unknowns are $\theta$ which measures the arm angle measured from the horizon and $\alpha$ which measures the stem angle measured from the east direction. The following trigonometric equations are used:

\begin{equation}
\label{eq:eq1}
a=L_{1}\sin\theta +L_{2}\cos\theta +L_{3}(\sin\alpha)(\sin\theta)-L_{4}
\end{equation}
\begin{equation}
\label{eq:eq2}
c=L_{3}\cos\alpha
\end{equation}
\begin{equation}
\label{eq:eq3}
d=L_{1}\cos\theta-L_{2}\sin\theta+L_{3}(\sin\alpha)(\cos\theta)
\end{equation}
\begin{equation}
\label{eq:eq5}
tan(\varepsilon)=\left(\frac{a\sin(\xi)}{c} \right)^2
\end{equation}
\begin{equation}
\label{eq:eq6}
\sin(\xi)=-\left(\frac{c}{d}\right)
\end{equation}

The arm angle $\theta$ can range from $45$ to $135$ degrees, whereas the stem angle $\alpha$ can range from $0$ to $180$ degrees. Combinations of $\theta$ and $\alpha$ with increments of $0.5$ are used to compute the related $\xi$ and $\varepsilon$ sun ray angles values. The combination leading to the smallest difference between those values and the ones obtained by the Jean Meeus algorithm is used.
 
\acknowledgments     
 
The research presented in this paper is funded by the Defence Science and Technology Agency (DSTA), Singapore. The authors would also like to thank Mr. Ong Kee Wang (NTU, Singapore) for building the initial design of WAHRSIS as a part of his final year project, and Mr. Alexandros Fragkiadakis (EPFL, Swizterland) for his contributions in the camera calibration as part of an internship at ADSC.


\bibliographystyle{spiebib}   

\end{document}